# Single-Layer Graphene Nearly 100% Covering an Entire Substrate

Mingsheng Xu[1,*], Daisuke Fujita[2,3], Keisuke Sagisaka[2], Eiichiro Watanabe[4] and Nobutaka Hanagata[4]

[1] International Centre for Young Scientists, National Institute for Materials Science, 1-2-1 Sengen, Tsukuba, Ibaraki 305-0047, Japan

[2] Advanced Nano Characterization Centre, National Institute for Materials Science, 1-2-1 Sengen, Tsukuba, Ibaraki 305-0047, Japan

[3] International Centre for Materials Nanoarchitectonics, National Institute for Materials Science, 1-2-1 Sengen, Tsukuba, Ibaraki 305-0047, Japan

[4] Nanotechnology Innovation Centre, National Institute for Materials Science, 1-2-1 Sengen, Tsukuba, Ibaraki 305-0047, Japan

*Corresponding author: XU.Mingsheng@nims.go.jp

**Graphene has recently attracted a great deal of interest in both academia and industry because of its unique electronic and optical properties[1,2], as well as its chemical, thermal, and mechanical properties. The superb characteristics of graphene make this material one of the most promising candidates for various applications, such as ultrafast electronic circuits[1] and photodetectors[2], clean and renewable energy[3], and rapid single-molecule DNA sequencing[4,5]. The electronic properties of the graphene system rely heavily on the number of graphene layers[6] and effects on the coupling with the underlying substrate. Graphene can be produced by mechanical exfoliation of graphite, solution approaches[7,8], thermal decomposition of $SiC$[9,10], and chemical vapor deposition/segregation on catalytic metals[11-17]. Despite significant progress in graphene synthesis, production with fine control over the**

**thickness of the film remains a considerable challenge. Here, we report on the synthesis of nearly 100% coverage of single-layer graphene on a Ni(111) surface with carbon atoms diffused from a highly orientated pyrolytic graphite (HOPG) substrate. Our results demonstrate how fine control of thickness and structure can be achieved by optimization of equilibrium processes of carbon diffusion from HOPG, segregation from Ni, and carbon diffusion at a Ni surface. Our method represents a significant step toward the scalable synthesis of graphene films with high structural qualities and finely controlled thicknesses and toward realizing the unique properties of graphene.**

Among the substrates used for graphene synthesis[10-21], nickel provides one of the smallest lattice mismatches with graphene, with a Ni(111) surface lattice constant of 0.249 nm compared to HOPG(0001) of 0.246 nm[12]. Thus, Ni(111) is one of the most promising catalytic metals for commensurate epitaxial growth of structurally homogeneous graphene. We deposited Ni film on a freshly cleaved HOPG(0001) substrate, and then treated the samples in a vacuum to form graphene layer on the Ni surface (Figures 1a and 1b) by controlling the cooling rate[22].

We used Auger electron spectroscopy (AES) to evaluate the uniformity and thickness of the graphene[9,12,17] on the Ni substrate because our graphene layer on the Ni substrate does not show any Raman characteristics between 1,000 and 3,500 $cm^{-1}$ when using 532 nm and 514 nm excitation lasers. Figure 2 shows a scanning electron microscopy (SEM) image with the corresponding Auger electron elemental maps (pixels: 512 × 512, and size: 26.7 μm × 26.7 μm) of the Ni MVV, C KLL, and Ni LMM of the graphene on Ni, providing information on the defects and film homogeneity on the Ni substrate with a high resolution. The AES spectra in Fig. 3 were obtained from the bright and gray regions of the sample. A typical SEM image revealed the presence of Ni grains with different color contrasts and steps, as well as dark hole-like defects and small very bright protrusions. In Figure 2, the small dark spots indicated with the arrow in the SEM image display the brightest feature in the C KLL map. Although the C KLL map (Fig. 2c) does display a slight color contrast in the bright and gray regions, the peak-to-peak magnitudes of C KLL (Fig. 3c) and Ni MVV (Fig. 3b) Auger electron spectra show negligible variation with sample regions, in contrast to considerable changes of the Ni LMM (Fig. 3d) electron intensity. Corresponding to the change in Ni LMM Auger electron intensity, the Ni LMM map (Fig. 2d) shows strong contrast in the bright and gray regions in the SEM image (Fig. 2a), which may indicate that the color contrast, *i.e.*, bright and gray grains in the SEM image, mainly stemmed from different electron emissions from the underlying Ni grains showing different crystalline orientations at an ~10 kV acceleration voltage. Together with larger area characterizations, the identical C KLL Auger electron intensity suggests a uniformly thick graphene film covering the bright and gray Ni grains.

To understand growth dynamics and to determine the thickness of the graphene film on Ni, the sample was sputtered by rastering a focused 1.0 keV Ar$^+$ ion beam over an area of 2 mm × 2 mm, and AES spectra were acquired as functions of sputter time. The C KLL (Fig. 3f) electron intensity was greatly decreased, while the intensities of Ni MVV (Fig. 3e) and Ni LMM electrons increased after the first 60 s of sputtering, and further changes along the same trend were observed with additional 30 s of sputtering. After a total of 90 s of sputtering, the intensities did not change further, but a carbon signal was still detected (Fig. 3f) that corresponds to carbon that remained in the Ni thin film, which was diffused from the HOPG substrate. According to these observations, our graphene formation is strikingly different from not only the growth mechanism of graphene on Cu foil[16] but also the graphene formation mechanism of other Ni-based methods[12-15]. The carbon atoms that form our graphene layer are controlled by the HOPG substrate by a diffusion process through the Ni template. Segregation of these carbon atoms from the Ni film, nucleation at the Ni grain boundaries and finally diffusion of carbon atoms on the Ni(111) surface further controls the formation of our graphene layer. These dynamic processes depend very much on the time and annealing temperature of the sample during the preparation process; parameters that together with the saturation solubility of carbon in Ni – significantly less than 0.1% in weight after annealing a Ni thin film[23] – are critical for the formation of the our graphene samples.

We used a standard attenuation model to estimate the thickness of the graphene film on the Ni substrate[24]. Because a low-energy electron is more sensitive to surface changes, the Ni MVV peak-to-peak magnitudes of the sample measured after 0 and 720 s of sputtering time (Fig. 3e) were used for calculation based on the following equation[25]:

$$I_{sub} = I_{sub,pure}\exp[-(nd_0)/\lambda\sin(\theta)], \qquad (1)$$

where $I_{sub}$ is the intensity of the Ni MVV Auger electron from the Ni substrate covered with graphene layers, $I_{sub,pure}$ is the intensity of the Ni MVV Auger electron from the sample after 700 s of sputtering, $n$ is the number of graphene layers, $d_0$ is the thickness of graphene (0.335 nm), $\theta$ is the electron take-off angle (42°) of the present Auger instrument, and $\lambda$ is the inelastic mean free path (IMFP) of the Auger electron for graphene systems. For the IMFP value at the electron energy of the Ni MVV transition, we used 4.45 Å, which we recently derived for graphenes[24]. By applying the IMFP value and the intensities of Ni MVV to the equation, the thickness of the graphene film was calculated as 3.05 Å on the Ni surface, suggesting a single-layer graphene film. This thickness value is smaller than the interlayer spacing (3.35 Å) of the C-C distance; however, it is reasonable, considering the interfacial orbital hybridization between graphene and

Ni[26-28].

To further confirm that our graphene sheet grown on Ni(111)/HOPG(0001) is a single-atomic layer of graphene, we transferred the graphene sheet from the Ni substrate on a silicon substrate coated with a ~200 nm SiO$_2$ layer, and then characterized it with Raman spectroscopy (Fig. 4). The Raman spectra show typical features of monolayer graphene[29]: a symmetric 2D band centered at ~2680 cm$^{-1}$ with a full width at half maximum of ~40 cm$^{-1}$, a G band centered at ~1580 cm$^{-1}$, and a G/2D intensity ratio of ~0.31. All these experimental evidences indicate we had synthesized a large-scale single-atomic graphene layer on Ni(111)/HOPG(0001). The Rama spectra also show a faint D band (~1350 cm$^{-1}$) associated with and almost negligible density of defects[29] that may be either inherent to the original graphene sheet or caused by the transfer[30] to the SiO$_2$/Si substrate.

Analyses of SEM images, Auger elemental maps, and AES spectra over the entire sample (2 cm by 2 cm) showed that the macroscopic defects induced by holes, very bright protrusions, and grain boundaries were less than 1% in area. Thus, including the defects, the single-layer graphene film covers more than 99% of the entire sample. It should be noted that such defects can be eliminated by optimizing growth conditions and carefully preparing the HOPG substrate. Furthermore, the sample size is easily scalable to larger dimensions, limited only by the size of the substrate and the growth chamber.

Our experiments provide evidence for the feasibility of synthesizing scalable single-layer epitaxy graphene covering 100% of a surface. We anticipate that the grown graphene will have distinct transport characteristics due to the almost nonexistent lattice mismatch in the system, *i.e.*, between HOPG(0001) and Ni(111) and between Ni and graphene, which is drastically different from graphenes on SiC, Ru, Pd, Pt, Ir, or Cu. Significant differences between the present system and other Ni-based graphene syntheses clearly lie in the process conditions, growth dynamics, and level of achievable thickness control. Other Ni-based growth (~1000 ºC) leads to graphene films with mixed numbers of layers[23]. Epitaxy growth of our system at a lower temperature (~650 ºC) produces uniform monolayer graphene in controlled modes of carbon diffusion from HOPG, solubility in Ni, and carbon diffusion at the Ni surface. The availability of such uniform-thickness graphene will provide an interesting system for the exploration of relevant properties of non-mechanical exfoliated graphene and for practical large-scale applications. Our approach opens a new venue for guiding the graphene production process, as graphene films with high crystalline quality and uniform thickness are essential to realize unique electron transport for graphene-based technologies.

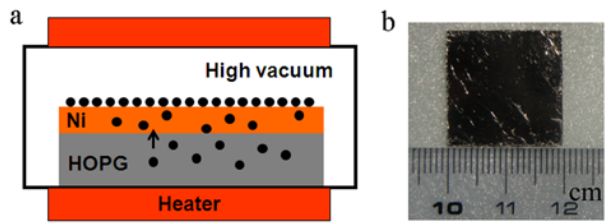

**Figure 1. Novel system for production of high-quality single-atomic graphene layer. a**, Schematic diagram of graphene growth system and formation mechanism. **b**, graphene on Ni(111)/HOPG(0001) (size: 2 cm × 2 cm).

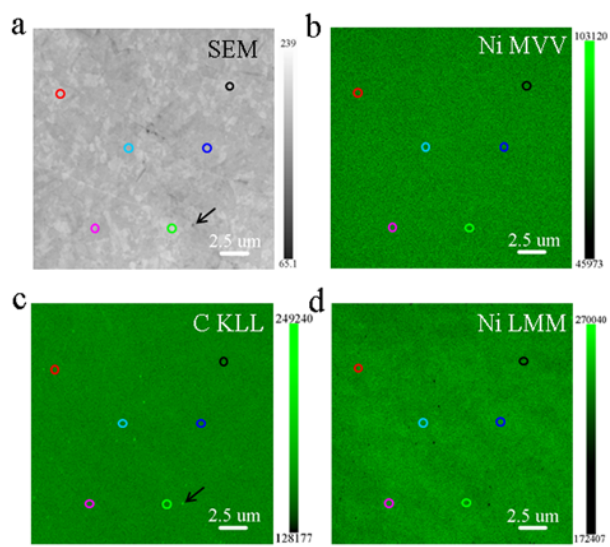

**Figure 2. AES characterization of graphene sheet grown on Ni(111)/HOPG(0001). a**, SEM image, showing brighter and darker grains, as well as small dark spots indicated by the arrow. **b**, Ni MVV Auger electron map. **c**, C KLL Auger electron map. **d**, Ni LMM Auger electron map.

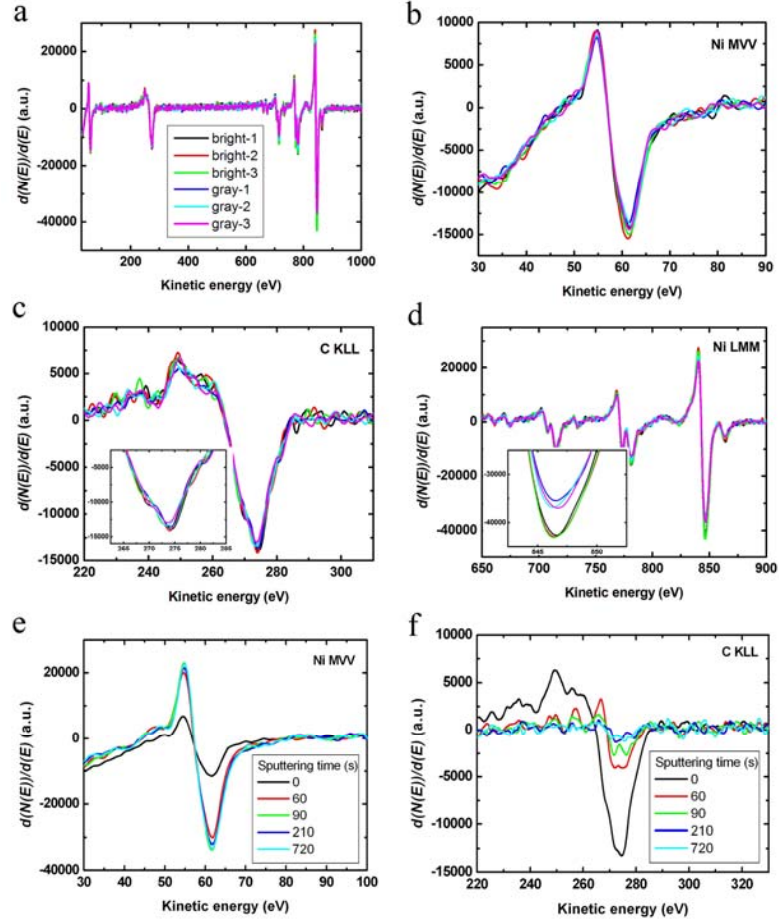

**Figure 3. AES spectra of graphene on Ni substrate.** (a) − (d) AES spectra of the graphene layer on Ni obtained from various bright and gray regions, as observed in the SEM images. (a) Survey spectra; (b) Ni MVV AES differential spectra; (c) C KLL AES differential spectra, showing no intensity change in the bright and gray regions; (d) Ni LMM AES differential spectra; (e) − (f) evolution of Auger electron intensity of graphene on Ni with sputtering time. (e) Ni MVV AES differential spectra; (f) C KLL AES differential spectra. After 90 s of sputtering, a carbon signal was still detected but did not change with further sputtering, indicating the carbon that remained in the Ni film.

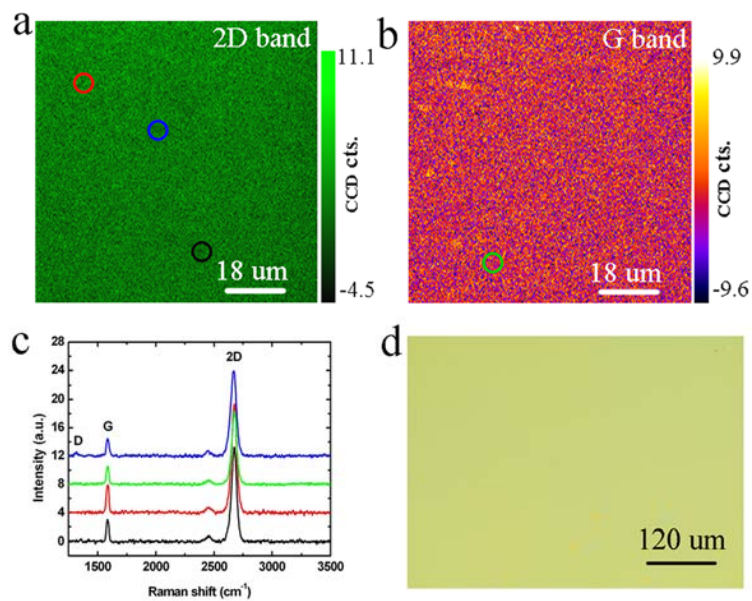

**Figure 4. Raman spectroscopy characterization of graphene sheet transferred on SiO$_2$/Si (200-nm-thick oxide layer), showing large-scale, single-atomic graphene layer. a**, Raman 2D band (~2680 cm$^{-1}$). **b**, Raman G band (~1580 cm$^{-1}$). **c**. Raman spectra from marked spots with corresponding colored squares as in **a**, **b**; and a defective D band (~1350 cm$^{-1}$) appeared in one of the spectra. **d**, Optical microscope image of the graphene, showing that it is optically uniform.